 \documentclass[prc,amsmath,amsfonts,showpacs,eqsecnum,floatfix,twocolumn]{revtex4}
 \usepackage{graphicx}  

\newcommand{\beq}{\begin{eqnarray}}
\newcommand{\eeq}{\end{eqnarray}}

\begin{document}

\title{Quark-quark interaction and quark matter in neutron stars}

\author{Y.\ Yamamoto$^{1}$}
\email{yamamoto@tsuru.ac.jp}
\author{N.\ Yasutake$^{2}$}
\author{Th.A.\ Rijken$^{3}$$^{1}$}
\affiliation{
$^{1}$RIKEN Nishina Center, 2-1 Hirosawa, Wako, 
Saitama 351-0198, Japan\\
$^{2}$Department of Physics, Chiba Institute of Technology, 2-1-1 Shibazono
Narashino, Chiba 275-0023, Japan\\
$^{3}$IMAPP, Radboud University, Nijmegen, The Netherlands
}

%

\begin{abstract}
Hyperon ($Y$) mixing in neutron-star matter brings about a remarkable softening 
of the equation of state (EoS) and the maximum mass is reduced to a value far
less than $2M_{\odot}$. One idea to avoid this ``hyperon puzzle in neutron stars" 
is to assume that the many-body repulsions work universally for every kind of baryons.
The other is to take into account the quark deconfinement phase transitions 
from a hadronic EoS to a sufficiently stiff quark-matter EoS.
In the present approach, both effects are handled in a common framework.
As well as the hadronic matter, the quark matter with the two-body quark-quark
interactions are treated within the Brueckner-Bethe-Goldstone theory 
beyond the mean field frameworks, where interaction parameters are 
based on the terrestrial data.
The derived mass-radius relations of neutron stars show that maximum masses 
reach over $2M_{\odot}$ even in the cases of including 
hadron-quark phase transitions, 
being consistent with the recent observations for maximum masses and radii
of neutron stars by the NICER measurements and the other multimessenger data.
\end{abstract}


\maketitle

\parindent 15 pt

\section{Introduction}

In studies of neutron stars, the fundamental role is played by
the equation of state (EoS) for dense nuclear matter.
The observed masses of neutron stars J1614$-$2230~\cite{Demorest10}, 
J0348+0432~\cite{Antoniadis13} and J0740+6620~\cite{Cromartie2020}
are given as $(1.97\pm0.04)M_{\odot}$, $(2.01\pm0.04)M_{\odot}$ and
$(2.17^{+0.11}_{-0.10})M_{\odot}$, respectively, being important 
conditions for the stiffness of the EoS of neutron-star matter.
In non-relativistic approaches, the stiff EoS giving the maximum 
mass of $2M_{\odot}$ can be derived from the existence of strongly 
repulsive effects such as three-nucleon repulsions 
in the high-density region \cite{APR98}. 

The hyperon ($Y$) mixing in neutron-star matter brings about 
a remarkable softening of the EoS and a maximum mass is reduced
to a value far less than $2M_{\odot}$.
The mechanism of EoS softening is understood as follows:
With increasing of baryon density toward centers of neutron stars,
chemical potentials of neutrons become high so that neutrons at
Fermi surfaces are changed to hyperons ($Y$) via strangeness 
non-conserving weak interactions overcoming rest masses of hyperons.
Then, it should be noted that naively such a mechanism of EoS softening 
works also for mixing of any exotic particles such as quarks into neutron matter.

One of the ideas to avoid this ``hyperon puzzle in neutron stars" is to assume
that the many-body repulsions work universally for every kind of baryons \cite{NYT}.
In Refs.\cite{YFYR14}\cite{YFYR16}\cite{YTTFYR17}, the multi-pomeron exchange
potential (MPP) was introduced as a model of universal repulsions among three 
and four baryons on the basis of the extended soft core (ESC) baryon-baryon 
interaction model developed by two of the authors (T.R. and Y.Y.) 
and M.M. Nagels~\cite{ESC16}.

Another solution for the hyperon puzzle has been suggested by taking
into account quark deconfinement phase transitions from a hadronic-matter 
EoS (H-EoS) to a sufficiently stiff quark-matter EoS (Q-EoS)
in the neutron-star interiors, namely by studying hybrid stars 
having quark matter in their cores \cite{Schaffner99} \cite{Baldo2006} \cite{Lastowiecki2012}
\cite{Shahrbaf1} \cite{Shahrbaf2} \cite{Maslov19} \cite{Xia19} 
\cite{Kojo2015} \cite{Baym2018} \cite{Otto2020}.
It is well known that repulsive effects in quark phases are needed
to result in massive neutron stars of $2M_{\odot}$.
In the Nambu-Jona-Lasinio (NJL) model, for instance, repulsions to stiffen 
EoSs are given by vector interactions \cite{Kunihiro} , strengths of which
are treated as phenomenological parameters to stiffen the EoSs.
Note that NJL models, including extended ones, are mainly based on 
mean-field approximations, in which two-body quark-quark interactions 
are not used explicitly. In spite of many works for hadron-quark phase 
transitions in neutron-star matter, there is not yet a unified theory 
of both the hadronic and quark phases.

In this work, our approach to hadron-quark phase transitions is
different from the usual methods in which the deconfined quark phases 
are treated in mean-field approximations. We handle here the quark matter 
with the two-body quark-quark ($QQ$) potentials derived as follows:
The meson-exchange quark-quark potentials are derived from the ESC
baryon-baryon ($BB$) potentials in the framework of the constituent quark model (CQM). 
The quark-quark-meson ($QQM$) vertices are defined such that, upon folding 
with the Gaussian ground-state baryonic quark wave functions, 
the $BB$ potentials are reproduced \cite{QQint}.
In this process the $QQM$ couplings are related to the $BBM$ couplings, 
and the extra interactions at the quark level necessary to achieve 
this connection are completely determined.
(Like in the ESC16 $BB$-potentials, relativistic effects are included in the 
$QQ$-potentials via the small components of the Dirac-spinors and a $1/M_Q$-expansion.)
The quark-quark instanton-exchange potential is derived from tuning the
baryon masses ($N, \Lambda, \Sigma, \Xi$), and $\Delta_{33}$ in the CQM.
Here, also the one-gluon-exchange (OGE) and the confining potential are included.

With use of these $QQ$ potentials together with the ESC $BB$-potentials,
baryonic matter and quark matter are treated in the common framework of 
the Brueckner-Bethe-Goldstone (BBG) theory, where the transitions 
between them are described in a reasonable way.
It should be emphasized here that our $QQ$ potentials are determined 
on the basis of the terrestrial data and do not include parameters only 
for the purpose of stiffening the quark-matter EoS.

Recently, the radius measurement has been performed for the most
massive neutron star PSR J0740+6620:
The two analyses have been done independently for the X-ray data
taken by the {\it Neutron Star Interior Composition Explorer} (NICER)
and the X-ray Multi-Mirror (XMM-Newton) observatory.
The radius and mass are $12.39^{+1.30}_{-0.98}$ km and
$2.072^{+0.067}_{-0.066}$ M$_\odot$ \cite{Riley2021} or
$13.7^{+2.6}_{-1.5}$ km ($68\%$) and $2.08 \pm 0.07$ M$_\odot$ \cite{Miller2021}.
The radius of a typical 1.4M$_\odot$ neutron star $R_{1.4M_\odot}$ has been 
estimated by combining the NICER measurements and the other multimessenger data
\cite{Raaij2021}\cite{Peter2021}.
In Ref.\cite{Raaij2021}, the two values of $R_{1.4M_\odot}=12.33^{+0.76}_{-0.81}$ km
and $R_{1.4M_\odot}=12.18^{+0.56}_{-0.79}$ km are obtained for the two different
high-density EoSs of a piecewise-polytrophic (PP) model and
a model based on the speed of sound, respectively. In Ref.\cite{Peter2021}, 
the estimated value is $R_{1.4M_\odot}=11.94^{+0.76}_{-0.87}$ km
at $90\%$ confidence. These values of radii are rather similar to each other.
On the other hand, when the implication of PREX-II for the neutron skin 
thickness of heavy nuclei are taken into account on the neutron-star EoS,
they obtain 13.33 km $< R_{1.4M_\odot} < 14.26$ km \cite{Brendan2021}.
Our obtained EoSs in this work are investigated in the light of these new data.

This paper is organized as follows:
In Sect.II, the hadronic-matter EoS (H-EoS) is recapitulated 
on the basis of our previous works.
In Sect.III, on the basis of realistic $QQ$ interaction models,
the BBG theory is applied to quark matter:
In III-A, the G-matrix framework is outlined for quark matter.
In III-B, our $QQ$ potentials are explained, which are composed of
the extended meson-exchange potential, the multi-pomeron potential,
the instanton potential and the one-gluon exchange potential.
In III-C, the $QQ$ G-matrix interactions in coordinate space are
parameterized as density-dependent interactions.
In Sect.IV, there are obtained the quark-matter EoSs (Q-EoS) 
and $MR$ diagrams of hybrid stars:  In IV-A, Q-EoSs are derived.
In IV-B, hadron-quark phase transitions in hybrid stars are
investigated on the basis of the obtained EoSs.
In IV-C, the $MR$ relations of hybrid stars are obtained
by solving the TOV equation.
The conclusion of this paper is given in Sect.V.

\section{Hadronic-matter EoS}

Here, the hadronic matter is defined exactly as
$\beta$-stable baryonic matter including leptons.
On the basis of the BBG theory,
the hadronic-matter EoS (H-EoS) is derived with use of the ESC 
baryon-baryon interaction model \cite{YFYR14}\cite{YFYR16}\cite{YTTFYR17}. 
Then, the EoS is stiff enough to assure the neutron-star masses of $2M_{\odot}$, 
if the strong three-nucleon repulsion is taken into account. 
However, the hyperon ($Y$) mixing results in remarkable softening 
of the EoS canceling this repulsive effect. In order to avoid 
this ``hyperon puzzle", it is assumed that the repulsions work 
universally for $Y\!N\!N$, $Y\!Y\!N$ $Y\!Y\!Y$ as well as for $N\!N\!N$.
In \cite{YFYR14}\cite{YFYR16}\cite{YTTFYR17}, such universal repulsions
are modeled as the multi-pomeron exchange potential (MPP). 
In Ref.\cite{YTTFYR17} they proposed three versions of MPP: MPa, MPa$^+$, MPb. 
MPa and MPa$^+$ (MPb) include the three- and four-body (only three-body) MPPs,
where mixings of $\Lambda$ and $\Sigma^-$ hyperons are taken into account.
The three-body part of MPa (MPa$^+$) is less repulsive than (equal to) that of MPb,
and the four-body parts of MPa and MPa$^+$ are equal to each other.
The EoSs for MPa and MPa$^+$ are stiffer than 
that for MPb, because of which radii of neutron stars obtained from 
the formers are larger than those from the latter.
Our ESC $BB$ interactions including MPb, MPa and MPa$^+$
are named as H1, H2, H3, for simplicity.
In addition, we introduce two versions H0 and H1' for comparative studies:
H0 is the nucleon-nucleon part of H1, being used in nuclear-matter EoSs with 
no hyperons. H1' is the $BB$ interaction H1 in which MPP works only 
among nucleons. In the case of H1', the remarkable softening of the EoS is 
brought about by hyperon mixing.

As shown later, neutron-star radii $R$ for masses lower than about $1.5M_\odot$ are 
determined by H-EoSs even in our $MR$ diagrams including hadron-quark transitions. 
For the H-EoSs derived from the above $BB$ interactions, 
the obtained values of radii at $1.4M_\odot$ ($R_{1.4M_\odot}$) are 
12.4 km (H1), 13.3 km (H2) and 13.6 km (H3).

\section{Quark-Quark interaction and quark matter}

\subsection{G-matrix framework}

The BBG theory is adopted for studies of quark matter on the basis of 
two-body $QQ$ potentials given in Ref.\cite{QQint}.
Here, correlations induced by $QQ$ potentials are renormalized into 
coordinate-space G-matrix interactions, being considered as effective 
$QQ$ interactions to derive the Q-EoS. 
In this stage to construct G-matrix interactions in quark matter,
color quantum numbers are not taken into account.

We start from the G-matrix equation for the quark pair $f_1 f_2$ in 
quark matter, where $f_1$ and $f_2$ denote flavor quantum numbers ($u,d,s$):
\begin{eqnarray}
G_{cc_0}=v_{cc_0} + \sum_{c'} 
 v_{cc'} {Q_{y'} \over \omega -\epsilon_{f'_1}-\epsilon_{f'_2} }
G_{c' c_0} 
\label{eq:GM1}
\end{eqnarray}
where $c$ denotes a relative state $(y, T, L, S, J)$ with $y=f_1f_2$, 
$S$ and $T$ being spin and isospin quantum numbers, respectively.
Orbital and total angular momenta are denoted by $L$ and $J$,
respectively, with ${\bf J}={\bf L}+{\bf S}$:
A two-quark state is specified by $^{2S+1}L_J$.
In Eq.~(\ref{eq:GM1}), $\omega$ gives the starting energy in 
the starting channel $c_0$.
The Pauli operator $Q_y$ acts on intermediate quark states with $y=f_1f_2$.
We adopt for simplicity the gap choice for the intermediate states
in the G-matrix equation, meaning that an intermediate energy
$\epsilon_{f}$ is replaced by a kinetic-energy operator.
The G-matrix equation~(\ref{eq:GM1}) is represented in the
coordinate space, whose solutions give rise to G-matrix elements.

The quark single particle (s.p.) energy $\epsilon_f$ 
in quark matter is given by
\begin{eqnarray}
\epsilon_f(k_f)={\hbar^2k_f^2 \over 2m_f} + U_f(k_f) 
\label{eq:GM2}
\end{eqnarray}
where $k_f$ is a $f$-quark momentum ($f=u,d,s$).
The potential energy $U_f$ is obtained self-consistently
in terms of the G-matrix as
\begin{eqnarray}
&& \hspace{-5mm} U_f(k_f) = 
 \sum_{|{\bf k}_{f'}|} \langle {\bf k}_f {\bf k}_{f'}
\mid G_{ff'}(\omega=\epsilon_f(k_f)+\epsilon_{f'}(k_{f'})) \mid
{\bf k}_f {\bf k}_{f'} \rangle
\nonumber\\
\label{eq:GM3}
\end{eqnarray}
where $(TLSJ)$ quantum numbers are implicit.
Then, the potential energy per particle 
$\langle U \rangle = \sum_{f} \langle U_f \rangle$
is obtained by averaging $U_f(k_{f})$ over $f$:
\begin{eqnarray}
\langle U \rangle=
\frac32\ \sum_{f} \omega_f \int_0^{k_F^f} \frac{d^3k_{f}}{(2\pi)^3}
\ U_f(k_{f}) 
\label{eq:GM4}
\end{eqnarray}
where $\omega_f=\rho_f/(\sum_{f'} \rho_{f'})$ with
a $f$-quark density $\rho_f$.
Making a partial wave reduction of Eq.~(\ref{eq:GM3}) with 
explicit use of $TLSJ$ quantum numbers, $U_f(k_{f})$ is 
represented as a sum of $U^{TLSJ}_f(k_{f})$ obtained from
G-matrix elements $G_{ff'}^{TLSJ}$.

\subsection{Quark-Quark interactions}

Our $QQ$ interaction is given by
\begin{eqnarray}
V_{QQ} &=& V_{EME}+V_{INS}+ V_{OGE}+V_{MPP} 
\end{eqnarray}
where $V_{EME}$, $V_{INS}$, $V_{OGE}$ and $V_{MPP}$ are the extended 
meson-exchange potential, the instanton exchange potential, the 
one-gluon exchange potential and the multi-pomeron potential, 
respectively \cite{QQint}. The included parameters in our $QQ$ potential
are chosen so as to be consistent with physical observables as much as possible.
The contributions of the confining potential ($V_{conf}$) to $V_{QQ}$
are minor in quark matter, being omitted in this work.

The $V_{EME}$ $QQ$ potential is derived from the ESC16 $BB$ potential \cite{ESC16} 
so that the $QQM$ couplings are related to the $BBM$ couplings
through folding procedures with Gaussian baryonic quark wave functions.
Then, the $V_{EME}$ $QQ$ potential is basically of the same functional expression
as the ESC16 $BB$ potential. The explicit expressions for $V_{EME}$ $QQ$ 
potentials are given in Ref.\cite{QQint}, 
In the ESC modeling, the strongly repulsive components in $BB$ potentials
are described mainly by vector-meson and pomeron exchanges between baryons.
It should be noted that this feature persists in the $V_{EME}$ $QQ$ potential,
which includes the strongly repulsive components originated from
vector-meson and pomeron exchanges between quarks.

Multi-pomeron exchanges are expected to work not only among baryons
but also among quarks, in which the baryon mass $M_B$ is replaced by
the quark mass $M_Q=M_B/3$ and the pomeron-baryon-baryon coupling constant 
$g_{PBB}$ is replaced by the pomeron-quark-quark coupling constant $g_{PQQ}$. 
In this work, the $QQ$ multi-pomeron potential $V_{MPP}$ is derived 
from the version MPa for the MPP among baryons.  

The included parameters included in $V_{INS}$ and $V_{OGE}$ are 
chosen so as to reproduce basic features of baryon mass spectra. 
The form of the one-gluon exchange potential is given as
\begin{equation} 
V_{OGE}(r)=\frac14\, ({\bf \lambda^C_1 \cdot \lambda^C_2})\,\alpha_S\,
V_{vector}(m_G;r) 
\end{equation}
where $\lambda^C_a$, $a=1,,,8$ are the Gell-Mann matrices in color 
SU(3) space and $V_{vector}(m_G;r)$ is the vector-type one boson 
exchange potential. Its explicit form is given by Eq.(E9a) in Ref.\cite{QQint}.
The strength of $V_{OGE}$ is determined by the quark-gluon coupling 
constant $\alpha_S$, being fixed as $\alpha_S=0.25$ in this work.
The gluon mass $m_G$ is taken as 420 MeV \cite{Hut95}.
In quark matter, $({\bf \lambda^C_1 \cdot \lambda^C_2})$=
$-8/3, +4/3, +4/3, -8/3$ in states of $(S,T)$=$(0,0), (0,1), (1,0), (1,1)$,
respectively.

The instanton potential $V_{INS}$ is based on the SU(3) generalization 
of the 't Hooft interaction for (u,d,s) quarks. In the configuration space,
with the addition of the Gaussian cut-off $\exp(-k^2/\Lambda_I^2)$,
the local instanton potential is given as \cite{QQint}
\begin{eqnarray}
&& V_{INS}(r)= -(4/3-\mbox{\boldmath $\lambda^F_1 \cdot \lambda^F_2$})\, G_I \,
\left(\frac{\Lambda_I}{2\sqrt{\pi}}\right)^3
\nonumber
\\
&&\times \left[1+\frac{\Lambda_I^2}{2m_Q^2} \left(3-\frac12 \Lambda_I^2 r^2 \right)
 \left(1-\frac13\mbox{\boldmath $\sigma_1 \cdot \sigma_2$}\right)
\right]
\nonumber
\\
&&\times \exp \left(-\frac14\Lambda_I^2 r^2\right) 
\label{eq:INS}
\end{eqnarray}
where $\lambda^F_a$, $a=1,,,8$ are the Gell-Mann matrices in flavor SU(3) space
and $m_Q$ is the quark mass.
In two-quark states, \mbox{\boldmath $\lambda$}$^F$ operators
are reduced to \mbox{\boldmath $\tau$} operators of isospin. 
The strength of $V_{INS}$ is determined by coupling constant $G_I$ and 
cut-off mass $\Lambda_I$. They are taken as $G_I=2.5$ GeV$^{-2}$ and 
$\Lambda_I=0.55$ GeV, being estimated from the $\pi-\rho$ mass splitting.

\begin{figure}[ht]
\begin{center}
\includegraphics*[width=8.6cm]{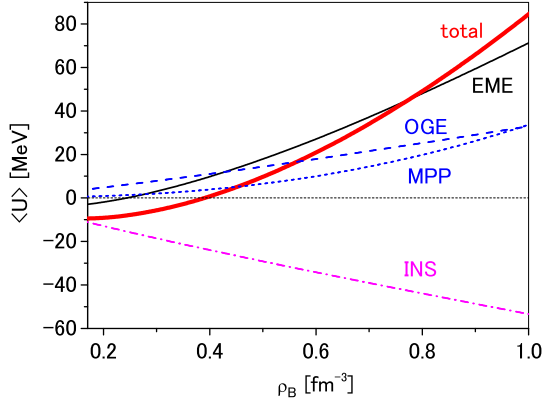}
\caption{Averaged single particle potentials $\langle U \rangle$
in quark matter as a function of the baryon number density 
$\rho_B=\frac13 \rho_Q$ in the case of $\rho_u=\rho_d=\rho_s$.
The solid, short-dashed, long-dashed and dot-dashed curves are
the contributions to $\langle U \rangle$ from $V_{EME}$, $V_{MPP}$,
$ V_{OGE}$ and $V_{INS}$, respectively. The bold-solid curve 
is obtained by $V_{EME}+V_{MPP}+V_{INS}+V_{OGE}$.
}
\label{Urho1}
\end{center}
\end{figure}

\begin{figure}[ht]
\begin{center}
\includegraphics*[width=8.6cm]{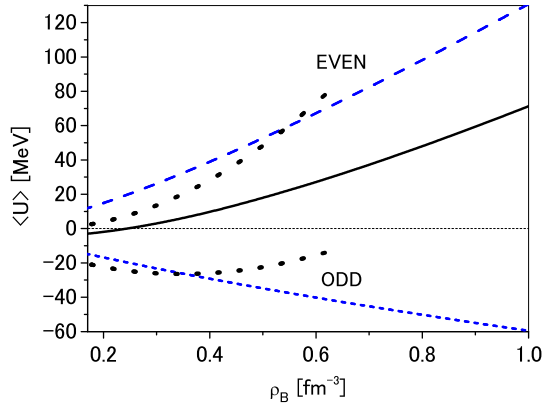}
\caption{Averaged single particle potentials $\langle U \rangle$
in quark matter as a function of the baryon number density 
$\rho_B=\frac13 \rho_Q$ in the case of $\rho_u=\rho_d=\rho_s$.
The solid curve is obtained from $V_{EME}$.
The even- and odd-state contributions 
$\langle U_{even} \rangle$ and $\langle U_{odd} \rangle$
are given by the dashed and short-dashed curves.
The dotted curves are the corresponding values of 
$\langle U_{even} \rangle$ and $\langle U_{odd} \rangle$ 
in neutron matter.}
\label{Urho2}
\end{center}
\end{figure}

In order to demonstrate the features of our $QQ$ interaction, we show 
the averaged s.p. potentials $\langle U \rangle$ given by Eq.~(\ref{eq:GM4})
as a function of the baryon number density $\rho_B=\frac13 \rho_Q$ 
in the case of $\rho_u=\rho_d=\rho_s$.
In Fig.~\ref{Urho1}, the solid, short-dashed, long-dashed and dot-dashed curves
are the contributions to $\langle U \rangle$ from $V_{EME}$, $V_{MPP}$,
$ V_{OGE}$ and $V_{INS}$, respectively. The bold-solid curve 
is obtained by the sum of $V_{EME}+V_{MPP}+V_{INS}+V_{OGE}$.
The strongly-repulsive nature of $\langle U(\rho_B) \rangle$ is the
key point in this work, which leads to the quark-matter EoS stiff 
enough to reproduce neutron-star masses over $2M_{\odot}$.
In the figure, the repulsive contribution of $V_{EME}$ is found to 
be essential the repulsive nature of $\langle U(\rho_B) \rangle$, 
where the repulsive contributions of $ V_{OGE}$ and $V_{MPP}$ are 
considerably canceled by the attractive contribution of $ V_{INS}$.
It is worthwhile to say that the repulsive components in $V_{EME}$
are from the vector-meson and pomeron exchanges. This feature persists
from the ESC $BB$ interaction model. 

In neutron matter, $\langle U \rangle$ includes repulsive contributions 
from the multi-pomeron potential MPP, being quite large in high density 
regions. The strengths of three- and four-body parts of MPP are proportional 
to $(g_{PBB})^3$ and $(g_{PBB})^4$, respectively, $g_{PBB}$ being the 
pomeron-baryon-baryon coupling constant.
In $QQ$ potentials, $g_{PBB}$ is replaced by the pomeron-quark-quark coupling 
constant $g_{PQQ}$. Because of the relation $g_{PQQ}=\frac13 g_{PBB}$, 
the strengths of three- and four-body parts of MPP among quarks are far smaller 
than those among baryons. Therefore, MPPs among quarks are not so remarkable 
in comparison with those among baryons.

The even- and odd-state contributions to $\langle U \rangle$ are denoted 
as $\langle U_{even} \rangle$ and $\langle U_{odd} \rangle$, respectively,
In Fig.~\ref{Urho2}, the solid curve shows $\langle U(\rho_B) \rangle$ 
obtained from $V_{EME}$, and $\langle U_{even}(\rho_B) \rangle$ and 
$\langle U_{odd}(\rho_B) \rangle$ are given by the dashed and short-dashed curves.
The dotted curves are the even- and odd-state contributions of
averaged neutron potentials $\langle U_{even} \rangle$
and $\langle U_{odd} \rangle$ in neutron matter.
The remarkable feature of $\langle U_{even} \rangle$ 
and $\langle U_{odd} \rangle$ given by $V_{EME}$ is that
they are attractive and repulsive, respectively. 
This feature of $\langle U_{even} \rangle$ and $\langle U_{odd} \rangle$
is similar to the corresponding one in neutron matter qualitatively.

When our $QQ$ potentials are used in quark-matter calculations,
it is reasonable to assume the constituent quark masses originated 
from the chiral symmetry breaking as the QCD non-perturbative effect.
Then, it is probable that the constituent quark masses in quark matter
become smaller than those in vacuum and move to current masses in
the high-density limit. 
At the mean-field (MF) level usually, the density-dependent quark masses
in matter have been derived from the MF-Lagrangian such as that of 
the NJL model. In the present approach, we introduce phenomenologically
the density-dependent quark mass
\begin{eqnarray}
M_Q^*(\rho_Q) = M_0/[1+\exp \{\gamma (\rho_Q-\rho_c\}] +m_0 +C
\label{mstar}
\end{eqnarray}
with $C=M_0-M_0/[1+\exp (-\gamma \rho_c)]$ assuring $M_Q^*(0) = M_0+m_0$,
where $\rho_Q$ is number density of quark matter, and $M_0$ and $m_0$ 
are taken as 300 (360) MeV and 5 (140) MeV for $u$ and $d$ ($s$) quarks.
Then, we have $M_Q^*(0)=$ 305 (500) MeV for $u$ and $d$ ($s$) quarks.  
The adjustable parameters $\rho_c$ and $\gamma$ are used to control 
mainly the onset densities of quark phases into hadronic phases.

Furthermore, because the quark mass reduction has to bring about 
an increase of the vacuum energy $B$, we assume simply
\begin{eqnarray}
B(\rho_Q)=M_Q^*(0) -M_Q^*(\rho_Q) \ .
\label{bag}
\end{eqnarray}

It is well known that there are three schemes for the 
density-dependent quark mass \cite{Blaschke20}:
(i) a constant quark mass, (ii) a linear density dependence
(Brown-Rho scaling \cite{Brown}), (iii) a density-dependence
within a higher-order NJL model \cite{Kashiwa} \cite{Benic}.
Eq.~(\ref{mstar}) includes these schemes, representing (i) for $\gamma=0$, 
(ii) for small values of $\gamma$ and  (iii) for large values of $\gamma$.
The parameter $\rho_c$ is chosen as $6\rho_0$ by referring 
to forms of (iii) derived from the higher-order NJL models. 

We define the following five sets with different values of $\gamma$
of $QQ$ interactions for deriving Q-EoSs.

\noindent
Q0 : $V_{EME}$ \ with $\gamma$=1.2

\noindent
Q1 (Q1e) : $V_{EME}+V_{INS}+V_{OGE}$ \ with $\gamma$=1.0 ($\gamma$=2.6)

\noindent
Q2 (Q2e) : $V_{EME}+V_{MPP}+V_{INS}+V_{OGE}$ \ with $\gamma$=1.6 ($\gamma$=2.2)

In the cases of Q0, Q1 and Q2, the values of $\gamma$ are chosen
so that the critical chemical potentials and densities for phase
transitions are as small as possible.
In the cases of Q1e and Q2e, they are chosen so that critical
densities are near crossing points of hadronic and quark
energy densities.

\begin{figure}[ht]
\begin{center}
\includegraphics*[width=8.6cm]{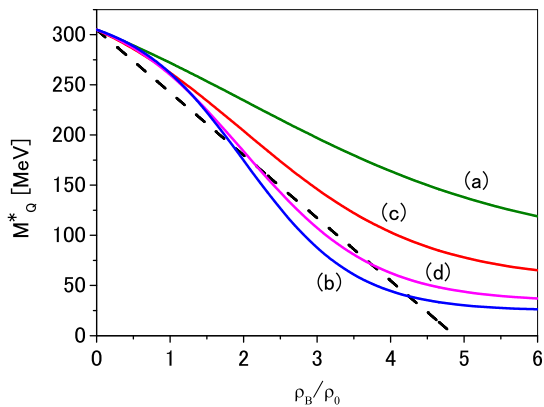}  
\caption{Quark mass as a function of the baryon number density $\rho_B$ for the
(a) Q1, (b) Q1e, (c) Q2 and (d) Q2e models. As a reference, also the Brown-Rho 
scaling is shown by the dashed line.}
\label{fig.quarkmass} 
\end{center}
\end{figure}

In Fig.~\ref{fig.quarkmass} the quark mass $M_Q^*(\rho_Q)$ ($Q=u,d$) as 
a function of the baryon number density $\rho_B=\rho_Q/3$ is plotted in
the cases of (a) Q1, (b) Q1e, (c) Q2 and (d) Q2e.
The density-dependent quark masses in these cases (especially Q1e and Q2e)
are found to be rather close to (ii) with the Brown-Rho scaling. 

It is quite important to use the density-dependent quark masses
together with our $QQ$ potentials. When constant quark masses are used, 
hadron-quark transitions derived from our $QQ$ potentials occur 
in density regions over $5\rho_0$ in hadronic matter giving 
$2M_{\odot}$ masses. In such a case, quark phases have no effect on 
masses and radii of neutron stars, even if they exist in inner cores.

For instance, when the baryon-baryon interaction H2 is used together 
with quark-quark interaction Q2, the combined set is denoted as H2+Q2.
Hereafter, combinations of $BB$ and $QQ$ interactions are expressed like this.

\subsection{Effective Quark-Quark interactions}

For applications to quark-matter calculations, we construct 
density-dependent effective local interactions ${\cal G}_{QQ}(\rho_Q;r)$
simulating G-matrices in coordinate space,
where $\rho_Q$ is number density of quark matter. 
We use here the method given in Ref.\cite{Yama10}.

The effective interactions are written as ${\cal G}_{QQ}={\cal G}_{EME}
+{\cal G}_{MPP}+{\cal G}_{INS}+{\cal G}_{OGE}$ approximately
corresponding to $V_{QQ}=V_{EME}+V_{MPP}+V_{INS}+V_{OGE}$. 
Though they can be obtained for each $(ff',T,L,S,J)$ state, 
for simplicity, the dependence on $L$ is approximated by 
that on parity $P$ and the dependence on $J$ is averaged:
Quantum numbers $TLSJ$ are reduced to $TSP$.
The respective interactions are represented in two- or one-range Gaussian forms,
and coefficients are adjusted so that s.p. potentials $U^{TSP}_f$
obtained from ${\cal G}_{ff'}^{TSP}$ simulate the original G-matrix results.
It is far easier to derive quark-matter EoSs with use of these 
density-dependent interactions ${\cal G}_{QQ}$ than derivations by 
G-matrix calculations with $V_{QQ}$.

The density-dependent effective interactions ${\cal G}_{EME}$ and ${\cal G}_{OGE}$
derived from $V_{EME}$ and $V_{OGE}$. respectively, are parameterized 
in a two-range Gaussian form as
\begin{eqnarray}
&& {\cal G}_{EME,OGE}(\rho,r) =  (a\rho^{\alpha}+b\rho^{\beta})\cdot \exp(-(r/0.8)^2)
\label{eq:EME}
\nonumber\\ 
	&& \hspace{2cm} + c\cdot \exp(-(r/1.6)^2) .
\end{eqnarray}
The parameter set $(a,\alpha,b,\beta,c)$ in Eq.(\ref{eq:EME}) is given 
for each $(y,T,S,P)$ state with $y=qq,qs,ss$ ($q=u,d$).
In Tables \ref{Geme} and \ref{Goge}, the values of parameters are
tabulated for ${\cal G}_{EME}$ and ${\cal G}_{OGE}$, respectively.

${\cal G}_{INS}$ derived from $V_{INS}$ is parameterized 
in an one-range Gaussian form as
\begin{eqnarray}
 {\cal G}_{INS}(\rho,r) =
  (a\rho^{\alpha}+b\rho^{\beta})\cdot \exp(-(r/0.6)^2) \ .
\label{eq:INS}
\end{eqnarray}
The parameter set $(a,\alpha,b,\beta,c)$ in Eq.(\ref{eq:INS}) is given 
for each $(y,T,S,P)$ state with $y=qq,qs$ ($q=u,d$).
The values of them are given in Table \ref{Gins}.

${\cal G}_{MPP}$ derived from $V_{MPP}$ is parameterized 
in an one-range Gaussian form as 
\begin{eqnarray}
 {\cal G}_{MPP}(\rho,r)= (a+b\rho^{\beta})\cdot \exp(-(r/1.3)^2) 
\label{eq:MPP}
\end{eqnarray}
being independent of $(y,T,S)$ and given only for $P$.
The values of parameters $(a,b,\beta)$ are given in Table \ref{Gmpp}. 

\begin{table}
\begin{center}
\caption{${\cal G}_{EME}(\rho,r)=(a\rho^{\alpha}+b\rho^{\beta})\cdot \exp(-(r/0.8)^2)
\\
\qquad \qquad + c\cdot \exp(-(r/1.6)^2)$.
$y=qq,qs,ss$ ($q=u,d$).
}
\label{Geme}
\vskip 0.2cm
\begin{tabular}{|c|c|ccccc|}\hline
$y$  & $T$ $S$ $P$  & $a$ & $\alpha$ & $b$ & $\beta$ & $c$ \\
\hline
$qq$ & 1 0 + & $-$3.520 &  $-1$ & $-$17.94 & 0 & $-$0.9978 \\
     & 0 1 + & $-$2.871 &  $-1$ & $-$30.59 & 0 & $-$0.8389 \\
     & 0 0 $-$ &    43.34 &  $-1$ &    192.8 & 0 &    3.896 \\
     & 1 1 $-$ &    6.621 &  $-1$ &    102.5 & 0 &    1.595 \\
\hline
$qs$ & 1/2 0 $+$ & $-$0.5716 &  $-1$ & $-$28.27 & 0 & $-$0.4530 \\
     & 1/2 1 $+$ & $-$0.6959 &  $-1$ & $-$24.58 & 0 & $-$0.1993 \\
     & 1/2 0 $-$ & $-$1.597 &  $-1$ &    149.0 & 0 &    1.568 \\
     & 1/2 1 $-$ &    1.183 &  $-1$ &    75.98 & 0 &    1.217 \\
\hline
$ss$ & 0 0 $+$ & $-$2.755 &  $-1$ & $-$26.37 & 0 & $-$0.1212 \\
     & 0 1 $-$ & $-$1.651 &  $-1$ &    51.06 & 0 &    0.3558 \\
\hline
\end{tabular}
\end{center}
\end{table}

\begin{table}
\begin{center}
\caption{${\cal G}_{OGE}(\rho,r)= (a\rho^{\alpha}+b\rho^{\beta})\cdot \exp(-(r/0.8)^2)
\\
\qquad \qquad +c\cdot \exp(-(r/1.6)^2)$. $y=qq,qs,ss$ ($q=u,d$).
}
\label{Goge}
\vskip 0.2cm
\begin{tabular}{|c|c|ccccc|}\hline
$y$   & $T$ $S$ $P$  & $a$ & $\alpha$ & $b$ & $\beta$ & $c$ \\
\hline
$qq$ & 1 0 + &    8.565 &  1  &    3.892 & 0.3742 &  0.5185 \\
     & 0 1 + &    7.543 &  1  &    1.977 & 0.6431 &  1.142 \\
     & 0 0 $-$ & -0.8959 &  1  &    9.982 & 0.2741 &  0.5027 \\
     & 1 1 $-$ &  8.094 &  1  &    11.64 & 0.2881 &  1.147 \\
\hline
$qs$ & 1/2 0 $+$ & $-$4.733 &  1  & $-$1.359 & 0.4161 & $-$0.2593 \\
     & 1/2 1 $+$ & $-$3.658 &  1  & $-$0.7316 & 0.6347 & $-$0.5709 \\
     & 1/2 0 $-$ & $-$1.282 &  1  & $-$3.141 & 0.3675 & $-$0.2514 \\
     & 1/2 1 $-$ & $-$5.645 &  1  & $-$3.831 & 0.3727 & $-$0.5736 \\
\hline
$ss$ & 0 0 $+$ &    10.48 &  1  &    1.478 & 0.5930 &  0.5185 \\
     & 0 1 $-$ &    11.23 &  1  &    7.743 & 0.3250 &  1.147 \\
\hline
\end{tabular}
\end{center}
\end{table}

\begin{table}
\begin{center}
\caption{${\cal G}_{INS}(\rho,r)=\ (a\rho^{\alpha}+b\rho^{\beta})\cdot \exp(-(r/0.6)^2)$. 
\\ $y=qq,qs$ ($q=u,d$).}
\label{Gins}
\vskip 0.2cm
\begin{tabular}{|c|c|cccc|}\hline
$y$   & $T$ $S$ $P$  & $a$ & $\alpha$ & $b$ & $\beta$  \\
\hline
$qq$ & 0 1 $+$ &   0.2132 & $-1$& $-$130.0 &   0   \\
     & 0 0 $-$ &    124.8 &  0  & $-$38.82 &  0.4227 \\
\hline
$qs$ & 1/2 0 $+$ & $-$1.504 & $-1$& $-$61.35 & 0.0705  \\
     & 1/2 1 $+$ & $-$0.1638 & $-1$& $-$63.47 &  0    \\
     & 1/2 0 $-$ &    59.03 &  0  & $-$17.06 & 0.4966  \\
     & 1/2 1 $-$ & $-$36.95 &  0  & $-$5.749 & 0.3192  \\
\hline
\end{tabular}
\end{center}
\end{table}

\begin{table}
\begin{center}
\caption{${\cal G}_{MPP}(\rho,r)= (a+b\rho^{\beta})\cdot \exp(-(r/1.3)^2)$.
}
\label{Gmpp}
\vskip 0.2cm
\begin{tabular}{|c|ccc|}\hline
  $P$  & $a$ & $b$ & $\beta$  \\
\hline
 $+$ & 0.3597  &  1.600 & 1.490  \\
 $-$ & 0.4338  &  2.618 & 1.384  \\
\hline
\end{tabular}
\end{center}
\end{table}

\section{EoS and $MR$ diagram of hybrid star}

\subsection{Derivation of quark-matter EoS}

Let us derive the EoS of quark matter composed of quarks with flavor $f=u,d,s$.
In this derivation, we use the density-dependent $QQ$ interactions
Eq.(\ref{eq:EME}), Eq.(\ref{eq:INS}), Eq.(\ref{eq:MPP})
based on the non-relativistic formalism. 
Relativistic expressions are used only for kinetic energies.

A single $f$ quark potential in quark matter composed of $f'$ quarks
is given by
\begin{eqnarray}
U_f(k)&=&\sum_{f'} U_{f}^{(f')}(k) 
 = \sum_{f'} \sum_{k'<k_F^{f'}} \langle kk'|{\cal G}_{ff',ff'}|kk'\rangle
\nonumber\\
\end{eqnarray}
with $f,f'=u, d, s$, where spin and isospin quantum numbers are implicit.
%
The quark energy density is given by
\begin{eqnarray}
\varepsilon_f&=&
 2N_c\sum_{f} \int_0^{k_F^f} \frac{d^3k}{(2\pi)^3}
\left\{\sqrt{\hbar^2 k^2+M_f^2}+\frac 12 U_f(k)\right\} 
	\nonumber\\ &&  + B(\rho_Q)
\label{eden} 
\end{eqnarray}
where $N_c=3$ is the number of quark colors.
The quark number density is given as $\rho_Q=\sum_f \rho_f$
with $\rho_f=N_c\frac{(k_F^f)^3}{3\pi^2}$. 
The chemical potential $\mu_f$ and pressure $P_Q$
are expressed as
\begin{eqnarray}
&&\mu_f = \frac{\partial \varepsilon_Q}{\partial \rho_f} \ , 
\label{chem} \\
&& P_Q = \rho_Q^2 \frac{\partial (\varepsilon_Q/\rho_Q)}{\partial \rho_Q}
 =\sum_f \mu_f \rho_f -\varepsilon_Q \ .
\label{press}
\end{eqnarray}

Here, we consider the EoS of $\beta$-stable quark matter 
composed of $u$, $d$, $s$, $e^-$. 
The equilibrium conditions are summarized as follows:

\noindent
(1) chemical equilibrium conditions,
\begin{eqnarray}
&& \mu_d = \mu_s = \mu_u+\mu_e 
\label{eq:c1}
\end{eqnarray}
\noindent
(2) charge neutrality,
\begin{eqnarray}
     0 = \frac13 (2\rho_u -\rho_d -\rho_s) -\rho_e 
\label{eq:c2}
\end{eqnarray}
\noindent
(3) baryon number conservation,
\begin{eqnarray}
\rho_B = \frac13 (\rho_u +\rho_d +\rho_s)= \frac13 \rho_Q
\label{eq:c3}
\end{eqnarray}

In the parabolic approximation, the following relation can be derived:
\begin{eqnarray}
\mu_e=\mu_d-\mu_u=4\beta E_{sym} 
\label{eq:c4}
\end{eqnarray}
where $x=\rho_u/(\rho_u +\rho_d)$ and $\beta=1-2x$.
$E_{sym}$ is the symmetric energy of $ud$ part. 

When the chemical potentials (\ref{chem})
are substituted into (\ref{eq:c1}),
the chemical equilibrium conditions 
are represented as equations for densities $\rho_u$,
$\rho_d$, $\rho_s$ and $\rho_e$.
Then, equations (\ref{eq:c1}) $-$ (\ref{eq:c3})
are solved iteratively, and densities and 
chemical potentials in equilibrium are obtained.
Finally, energy densities (\ref{eden}) and pressures
(\ref{press}) can be calculated.

\begin{figure}[ht]
\begin{center}
\includegraphics*[width=8.6cm,height=7.5cm]{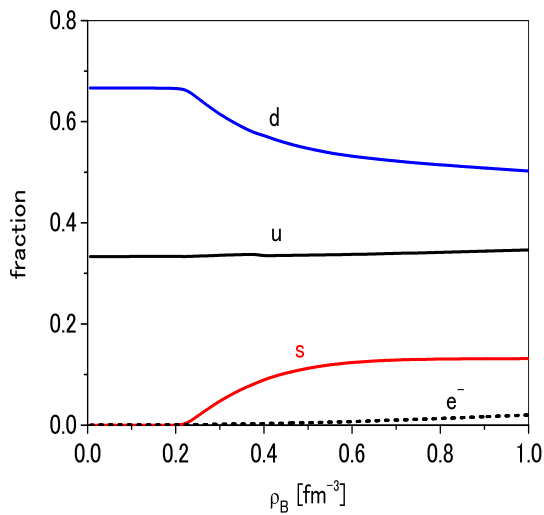}
\caption{The number fractions of quarks and electrons
in $\beta$-stable quark matter as a function of the
baryon density $\rho_B$ in the case of using Q2.
The fractions of $u$, $d$ and $s$ quarks are given by
solid curves, and that of electrons $e^-$ are by dashed curve.}
\label{comp}
\end{center}
\end{figure}

An example of solution is demonstrated in Fig.~\ref{comp}: 
The number fractions of quarks and electrons in $\beta$-stable 
quark matter are plotted as a function of the baryon density 
$\rho_B$ in the case of using Q2, where solid (dashed) curves 
are for $u$, $d$ and $s$ quarks (electrons).
In the figue, the electron fractions are not visible below 
the $s$-quark onset. The reason of such small values are because 
the symmetry energies $E_{sym}$ in Eq.(\ref{eq:c4}) are not so 
large in the case of our $QQ$ interactions.

\subsection{Phase transition from hadronic matter to quark matter}

\begin{figure}[ht]
\begin{center}
\includegraphics*[width=8.6cm,height=7.5cm]{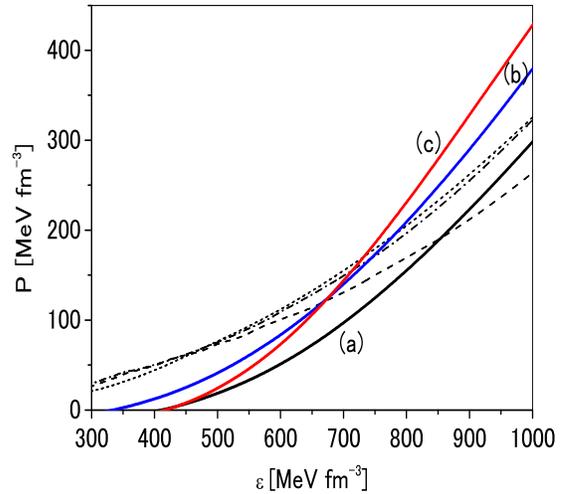}
\caption{Pressures of hadronic matter and quark matter as a function of 
the energy density $\epsilon$. Short-dashed, long-dashed and dot-dashed
curves are for hadronic matter for H1, H2 and H3, respectively.
Solid curves are for quark matter obtained from (a) Q0, (b) Q1 and (c) Q2.
}
\label{Peden}
\end{center}
\end{figure}

The EoSs are shown in Fig.~\ref{Peden}, where pressures of quark matter
are given as a function of the energy density $\epsilon$ and compared 
to those of hadronic matter.
Steeper slopes of curves correspond to stiffer EoSs:
The Q-EoSs are stiffer than the H-EoSs, and the EoSs for (b) Q1 and 
(c) Q2 are stiffer than that for (a) Q0 owing to the repulsive contributions 
of $V_{OGE}$ and $V_{MPP}$. As shown later, these features are clearly 
reflected in the $MR$ curves of hybrid stars.

In order to construct the hybrid EoS including a transition from 
hadronic phase and quark phase, 
we use the replacement interpolation method \cite{RIM} \cite{Shahrbaf2}, 
being a simple modification of the Maxwell and the Glendenning (Gibbs) 
constructions \cite{Glendenning}. 
In our actual calculations, we follow the interpolation formula 
given in Ref.\cite{Shahrbaf2}.
Then, interpolated regions can be considered as mixed phases.
Both of H-EoSs and Q-EoSs are assumed to fulfill separately 
the charge-neutrality and $\beta$-equilibrium conditions.
The EoSs of hadronic and quark phases and that of mixed phase are
described with the relations between pressures and chemical potentials
$P_H(\mu)$, $P_Q(\mu)$ and $P_M(\mu)$, respectively.
The critical chemical potential $\mu_c$ for the transition
from the hadronic phase to the quark phase is 
obtained from the Maxwell condition 
\begin{eqnarray}
P_Q(\mu_c)=P_H(\mu_c)=P_c  \ .
\end{eqnarray}
The pressure of the mixed phase is represented by a polynomial ansatz
\begin{eqnarray}
P_M(\mu)=\sum^N_{q=1} \alpha_q (\mu-\mu_c)^q +P_c+\Delta P 
\label{eq:PM}
\end{eqnarray}
where the pressure shift $\Delta P$ at $\mu_c$ is treated as
a free parameter. 
The pressure of the mixed phase at $\mu_c$ is determined by
$P_M(\mu_c)=P_c+\Delta P= (1+\Delta_P)P_c$ with 
$\Delta_P=\Delta P/P_c$. 
Then, the matching chemical potential $\mu_H$ ($\mu_Q$) of $P_M(\mu)$ 
to $P_H(\mu)$ ($P_Q(\mu)$) can be obtained from the continuity condition.
The corresponding matching densities $\rho_H$ and $\rho_Q$ are obtained
with use of $\rho(\mu)=dP(\mu)/d\mu$. 
The finite values of $\Delta_P=0.05 - 0.07$ corresponds to 
the Glendenning construction \cite{Shahrbaf2}. 
We choose $\Delta_P=0.07$ in this work.

\begin{figure*}[ht]
\begin{center}
\includegraphics*[width=6in,height=3in]{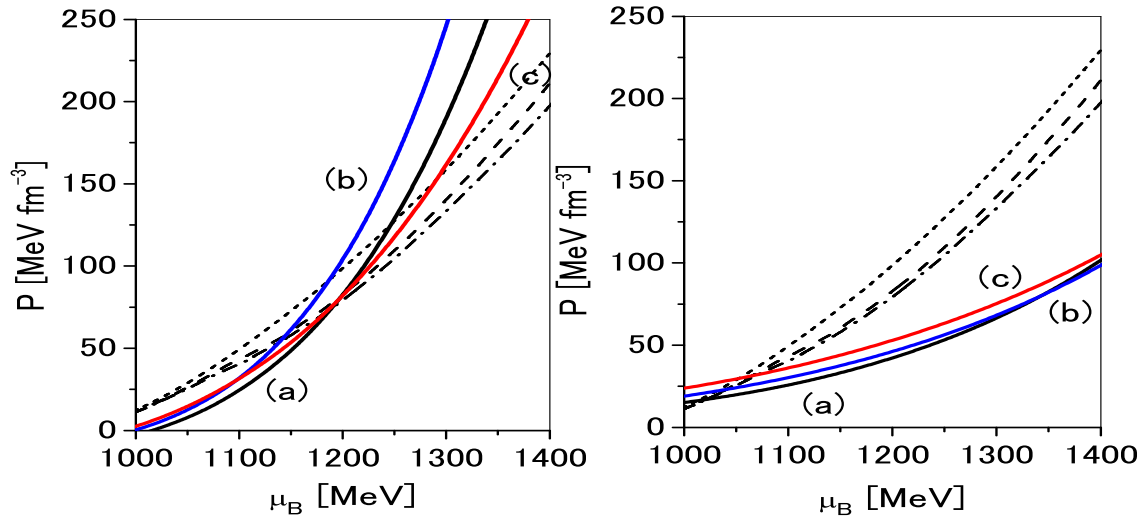}
\caption{Pressures as a function of the chemical potential $\mu_B$.
Short-dashed, long-dashed and dot-dashed curves are 
pressures of hadronic matter for H1, H2 and H3, respectively.
In the left panel, solid curves are pressures of quark matter obtained 
from (a) Q0, (b) Q1 and (c) Q2. In the right panel, they are obtained 
by using constant quark masses without density dependences 
Eq.~(\ref{mstar}) and vacuum energies Eq.~(\ref{bag}).
}
\label{Pmu}
\end{center}
\end{figure*}

In Fig.~\ref{Pmu}, pressures are drawn as a function of
the chemical potential $\mu_B$, where short-dashed, 
long-dashed and dot-dashed curves are pressures of 
hadronic matter for H1, H2 and H3, respectively.
In the left panel, solid curves are pressures of quark matter 
obtained from (a) Q0, (b) Q1 and (c) Q2. The crossing of the hadronic 
and the quark-matter curves is considered to be a condition for 
phase transition to occur.
The values of $P$ at crossing points give the critical pressures
$P_c$ for phase transitions. The hadronic and quark-matter curves 
are connected smoothly by Eq.~(\ref{eq:PM}).
Then, the effective-mass parameter $\gamma$ in Eq.~(\ref{mstar}) 
is adjusted so that cross points appear at similar values of 
$\mu_B \sim 1200$ MeV.
In the right panel, on the otherhand, solid curves are obtained
from Q0, Q1 and Q2 by using constant quark masses $M_Q^*(\rho_Q=0)$ 
without density dependences Eq.~(\ref{mstar}) and vacuum energies 
Eq.~(\ref{bag}). It is found that there is no crossing point in 
this region of $\mu_B$. Thus, the density-dependent quark mass 
plays a decisive role in the occurrence of phase transition.

\begin{figure}[ht]
\begin{center}
\includegraphics*[width=8.6cm,height=7.5cm]{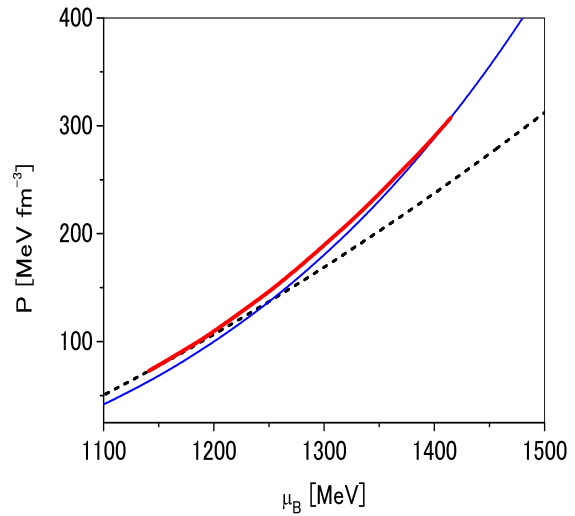}
\caption{Pressures as a function of the chemical potential $\mu_B$
in the transition region. The short-dashed curve is obtained 
by the H-EoS for H2 and the solid curve is by the Q-EoS for Q2. 
The bold-solid curve is the interpolated one.
}
\label{Mixed1}
\end{center}
\end{figure}

In Fig.~\ref{Mixed1}, pressures are given as a function of 
the chemical potential $\mu_B$ in the transition region.
The short-dashed curve is obtained by the H-EoS for H2 and 
the solid curve is by the Q-EoS for Q2. 
The bold-solid curve is obtained by the interpolation method.

\begin{table}
\begin{center}
\caption{Pressures $P_c$ at critical chemical potentials 
$\mu_c$ in phase transitions from the hadronic phases for H1, H2 
and H3 to the quark-matter phases for Q0, Q1 and Q2. Values of
$\mu_H$ ($\mu_Q$) are chemical potentials at matching points
between mixed phases and hadron (quark) phases. 
} 
\label{match1}
\vskip 0.2cm
\begin{tabular}{|l|cccc|}
\hline
 & $P_c$  & $\mu_c$  & $\mu_H$ & $\mu_Q$ \\
 & MeV/fm$^3$ & MeV & MeV & MeV  \\ 
\hline
H1+Q0 & 125.7 &  1241  & 1186   & 1373 \\
H1+Q1 & 92.55 &  1183  & 1141   & 1277 \\
H1+Q2 & 110.4 &  1215  & 1095   & 1368 \\
\hline
H2+Q0 & 139.6 &  1254  & 1199   & 1386  \\
H2+Q1 & 102.3 &  1193  & 1149   & 1282  \\
H2+Q2 & 138.2 &  1252  & 1141   & 1415  \\
H2+Q1e& 132.9 &  1189  & 1243   & 1446  \\
H2+Q2e& 209.6 &  1360  & 1261   & 1600  \\
\hline
H3+Q0 & 136.0 &  1251  & 1198   & 1382  \\
H3+Q1 & 101.2 &  1191  & 1148   & 1279  \\
H3+Q2 & 131.6 &  1243  & 1142   & 1412  \\
\hline
\end{tabular}
\end{center}
\end{table}

\begin{table}
\begin{center}
\caption{
Critical densities (fm$^{-3}$) of phase transitions:
$\rho_H$ and $\rho_Q$ are densities at matching points
in phase transitions from the hadronic phases for H1, H2 
and H3 to the quark-matter phases for Q0, Q1, Q2, Q1e and Q2e.
$\rho^c_H$ and $\rho^c_Q$ are critical densities for the 
Maxwell construction defined by $P_H(\rho^c_H)=P_Q(\rho^c_Q)=P_c$.
Values of $\rho_E$ are densities at crossing points of
energy densities $\epsilon_H(\rho)$ and $\epsilon_Q(\rho)$.
There is no crossing point in the case of H1+Q2.
} 
\label{match2}
\vskip 0.2cm
\begin{tabular}{|l|cc|cc|c|}
\hline
 &  $\rho_H$ & $\rho_Q$ &$\rho^c_H$ & $\rho^c_Q$  & $\rho_E$  \\
\hline
H1+Q0 & 0.566 &0.904 &0.661  &0.673  &0.784  \\
H1+Q1 & 0.490 &0.703 &0.574  &0.544  &0.918  \\
H1+Q2 & 0.407 &0.721 &0.623  &0.584  &  ---  \\
\hline
H2+Q0 & 0.521 &0.930 &0.664  &0.694  &0.702  \\
H2+Q1 & 0.446 &0.712 &0.573  &0.561  &0.753  \\
H2+Q2 & 0.433 &0.776 &0.661  &0.620  &0.716  \\
H2+Q1e& 0.506 & 1.02 &0.650  &0.707  &0.643  \\
H2+Q2e& 0.608 &0.987 &0.790  &0.722  &0.695  \\
\hline
H3+Q0 & 0.482 &0.922 &0.616  &0.689  &0.659  \\
H3+Q1 & 0.416 &0.706 &0.568  &0.559  &0.692  \\
H3+Q2 & 0.407 &0.772 &0.608  &0.612  &0.660  \\
\hline
\end{tabular}
\end{center}
\end{table}

Our phase transition is specified by the pressures $P_c$ at
critical chemical potentials $\mu_c$ and boundary values of 
chemical potentials and densities for mixed phases.
They are shown in the cases of phase transitions from the 
H-EoS for H1, H2 and H3 to the Q-EoSs for Q0, Q1, Q2, Q1e and Q2e.
In Table~\ref{match1},
the chemical potentials at matching points are given by 
values of $\mu_H$ and $\mu_Q$.
In Table~\ref{match2}, 
$\rho_H$ and $\rho_Q$ are densities at matching points
in phase transitions, 
and $\rho^c_H$ and $\rho^c_Q$ are critical densities
defined by the $P_H(\rho^c_H)=P_Q(\rho^c_Q)=P_c$ 
in the case of $\Delta_P=0$.
It is reasonable that that the values of $\rho^c_H$ and $\rho^c_Q$ are 
between the values of $\rho_H$ and $\rho_Q$.
The Maxwell construction is conditioned by $\rho^c_H < \rho^c_Q$.
As found in Table~\ref{match2}, however, the values of $\rho^c_H$ are larger 
than those of $\rho^c_Q$ in some cases, meaning that first-order phase transitions 
do not appear in $\Delta_P=0$ limits. 

The values of $\rho_E$ are densities at crossing points of
energy densities $\epsilon_H(\rho)$ and $\epsilon_Q(\rho)$.
In the case of H2+Q1e (H2+Q2e), the value of $\rho_E$ is between
(smaller than) $\rho^c_H$ and $\rho^c_Q$.   

\begin{figure}[ht]
\begin{center}
\includegraphics*[width=8.6cm,height=7.5cm]{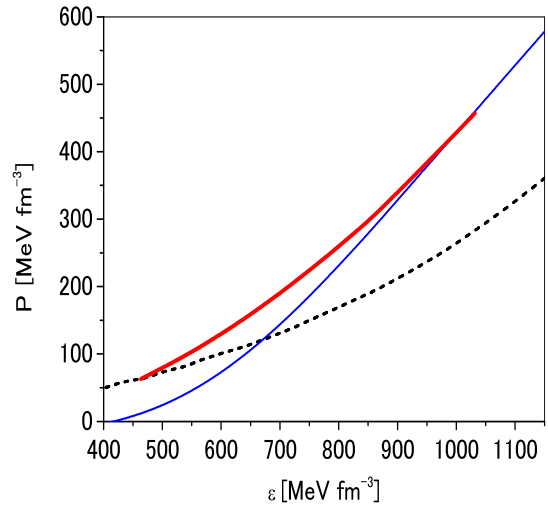}
\caption{Pressures as a function of as a function of the energy density $\epsilon$ 
in the region of phase transitions. The short-dashed curves are obtained by 
the H-EoS for H2 and the solid curve is by the Q-EoS for Q2. 
The bold-solid curve is the interpolated ones in the mixed phase.
}
\label{Mixed2}
\end{center}
\end{figure}

In Fig.~\ref{Mixed2},  pressures are given as a function 
of energy density $\epsilon$ in the transition region.
The short-dashed curve is obtained by the H-EoS for H2 and 
the solid curve is by the Q-EoS for Q2. 
The bold-solid curve is pressure in the interpolated region.

It is worthwhile to point out that our hybrid-EoSs are consistent with
the picture of hadron-quark continuity \cite{Kojo2015}\cite{Baym2018}.
In these references, the interpolated pressures are given in the density region of 
$2<\rho_B/\rho_0<(4 - 7)$, where quark degrees of freedom gradually emerge. 
Correspondingly, our mixed phases are given in the region of 
$(2.4 - 3.3)<\rho_B/\rho_0<(4.1 - 6.0)$,
as found in Table~\ref{match2}.

\begin{figure}[ht]
\begin{center}
\includegraphics*[width=8.6cm,height=7.5cm]{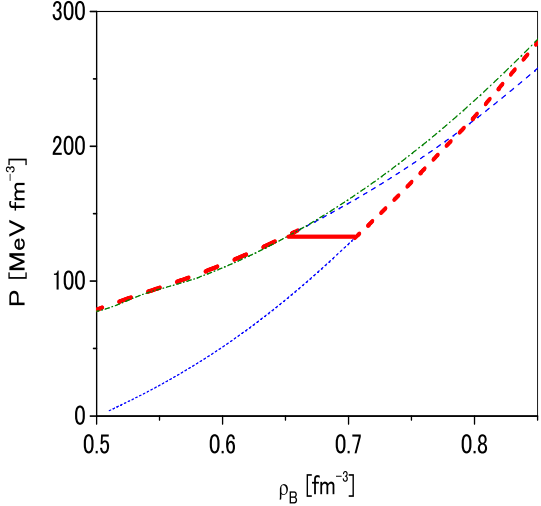}
\caption{Pressure $P$ as a function of density $\rho_B$ in the case of H2+Q1e, 
where the dashed (short-dashed) curves are for hadronic (quark) matter. 
The horizontal solid line shows the range of the hadron-quark mixed phases
in case of $\Delta_P=0$. The dot-dashed curve is in the case of $\Delta_P=0.07$.
}
\label{MRdel}
\end{center}
\end{figure}

Let us demonstrate that the Maxwell construction appears 
in the the $\Delta_P=0$ limit. In Fig.~\ref{MRdel}, we show pressure $P$ 
as a function of density $\rho_B$ 
in the case of using H2+Q1e, where the dashed (short-dashed) curves are 
for the hadronic (quark) matter. The horizontal solid lines show the 
range of the hadron-quark mixed phase.
It is well known that the Maxwell construction is specified by
the horizontal line in the $P-\rho$ diagram, where
the density values at ends of horizontal and vertical lines 
are given by $\rho^c_H$ and $\rho^c_Q$.
The difference of the curve for $\Delta_P=0$ from the dot-dashed curve 
for $\Delta_P=0.07$ is found to be small, the appearance of which is 
seen in the corresponding $MR$ curves as shown later.
Not only in the case of He+Q1e, there appear the similar curves
specifying the Maxwell construction in the cases of 
$\rho^c_H < \rho^c_Q$ in Table~\ref{match2}.

Our hybrid-star EoS is composed of H-EoS and Q-EoS, being combined 
by the interpolation formula including the parameter $\Delta_P$.
The $MR$ relations of hybrid stars can be obtained by solving the 
Tolmann-Oppenheimer-Volkoff (TOV) equation, where our hybrid EoSs 
are connected smoothly to the crust EoS \cite{Baym1}\cite{Baym2}
in the low-density side.

\begin{figure}[ht]
\begin{center}
\includegraphics*[width=8.6cm,height=5cm]{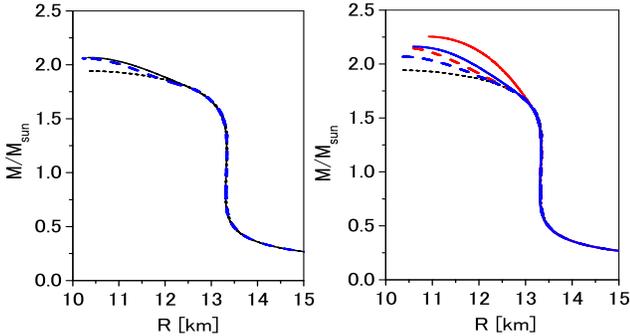}
\caption{Hybrid-star masses as a function of radius $R$,
where the short-dashed curves are obtained by the H-EoS for H2.
In the left panel, the solid and dashed curves are obtained by
H2+Q1e in cases of $\Delta_P=0.07$ and $\Delta_P=0$, respectively.
In the right panel,
the upper (lower) solid curves are obtained by H2+Q2 (H2+Q1),
and the upper (lower) dashed curves are by H2+Q2e (H2+Q1e).
}
\label{MRgam}
\end{center}
\end{figure}

In Fig.~\ref{MRgam}, hybrid-star masses are shown as a function of 
radius $R$ in the cases of using Q1e or Q2e (Q1 or Q2), where
the short-dashed curves are obtained by the H-EoS for H2.

In the left panel, the solid and dashed curves are obtained by H2+Q1e
in cases of $\Delta_P=0.07$ and $\Delta_P=0$ (Maxwell construction), 
respectively. The slight reduction of the latter compared to the former
is due to the difference between the $\Delta_P=0.07$ and $\Delta_P=0$
curves in in Fig.\ref{MRdel}.

In the right panel, the upper (lower) solid curves are obtained by 
H2+Q2 (H2+Q1) in the case of $\Delta_P=0.07$, 
and the upper (lower) dashed curves are by H2+Q2e (H2+Q1e).
The maximum masses are $2.15M_\odot$ ($2.25M_\odot$) for H2+Q2e (H2+Q2),
and $2.07M_\odot$ ($2.16M_\odot$) for H2+Q1e (H2+Q1).
The quark-phase onset values of the central baryon densities 
are 0.54 fm$^{-3}$ (0.48 fm$^{-3}$) in the case of H2+Q1e (H2+Q1),
and 0.65 fm$^{-3}$ (0.46 fm$^{-3}$) in the case of H2+Q2e (H2+Q2).
Thus, Q2e (Q1e) leads to the larger onset density
and the smaller maximum mass than Q2 (Q1):
It is a general trend that maximum masses become smaller
as onset densities of quark phases become larger.

In our approach, there is no clear criteria to decide which of Q2 (Q1) 
and Q2e (Q1e) is more appropriate. In the following section,
we use Q1 and Q2 because they seem to be more suitable than
Q1e and Q2e in the light of the recent observations for the maximum masses.

\subsection{$MR$ diagrams of hybrid stars}

\begin{figure*}[ht]
\begin{center}
\includegraphics*[width=6in,height=3in]{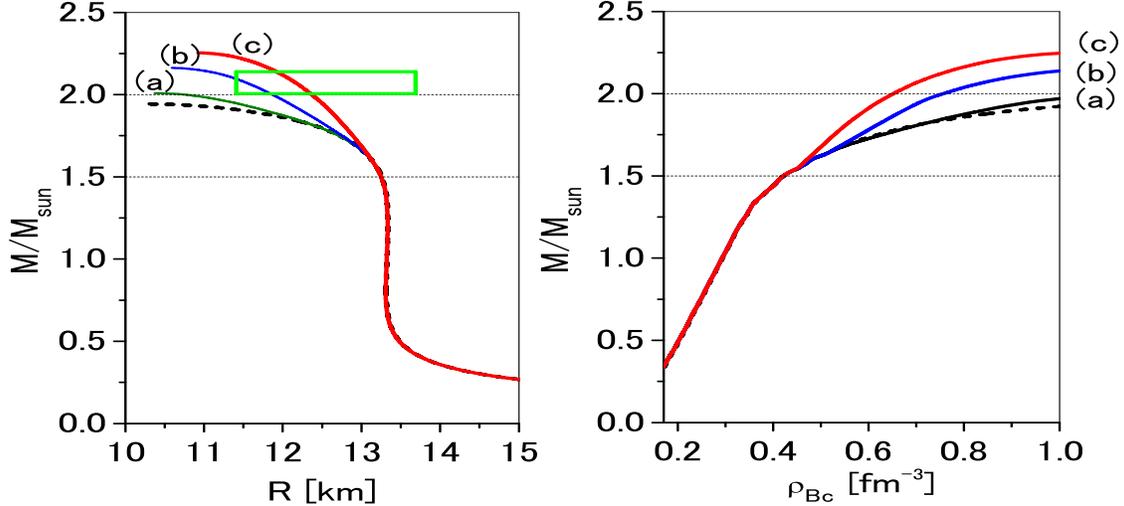}
\caption{Hybrid-star masses as a function of radius $R$ (left panel)
and central density $\rho_{Bc}$ (right panel).
The solid curves are obtained by the Q-EoSs for (a) Q0, (b) Q1 and (c) Q2.
The short-dashed curves are by the H-EoS for H2.
The deviations from the latter to the formers are by phase transitions 
from hadronic-matter to quark-matter. The rectangle indicates
the region of mass $2.072^{+0.067}_{-0.066}$M$_\odot$ and radius
$12.39^{+1.30}_{-0.98}$ km  \cite{Riley2021} for PSR J0740+6620.}
\label{MR1}
\end{center}
\end{figure*}

\begin{figure*}[ht]
\begin{center}
\includegraphics*[width=6in,height=3in]{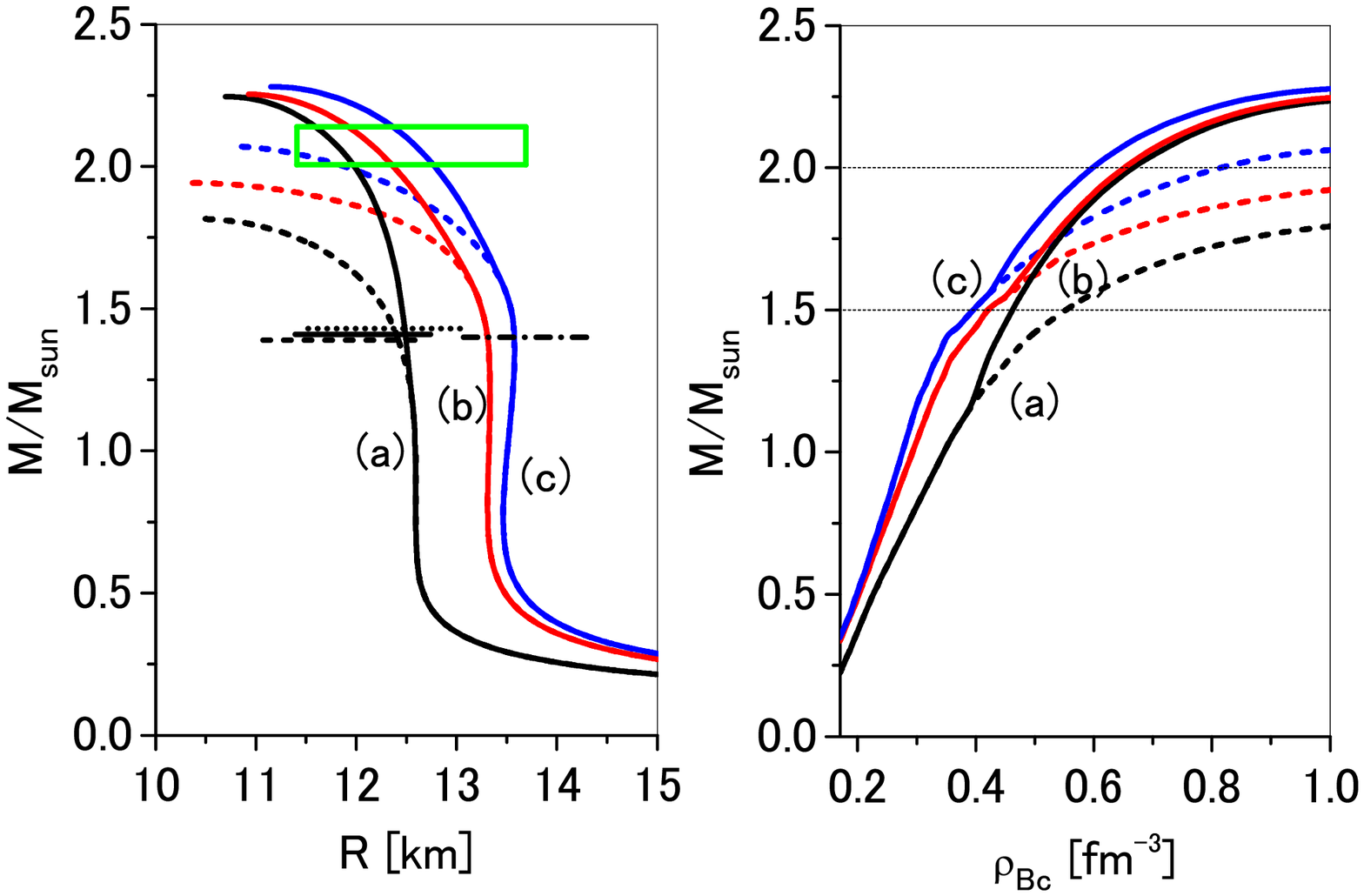}
\caption{Hybrid-star masses as a function of radius $R$ (left panel)
and central density $\rho_{Bc}$ (right panel),
where the Q-EoS for Q2 is used.
Short-dashed, long-dashed and dot-dashed curves are obtained with
H-EoSs for (a) H1, (b) H2 and (c) H3, respectively.
Solid curves show the deviations by transitions from hadronic-matter 
to quark-matter phases. In the left panel, the rectangle indicates
the region of mass $2.072^{+0.067}_{-0.066}$M$_\odot$ and radius
$12.39^{+1.30}_{-0.98}$ km  \cite{Riley2021}.
Dotted and solid line segments indicate $R_{1.4M_\odot}=12.33^{+0.76}_{-0.81}$ km
(PP model) and $R_{1.4M_\odot}=12.18^{+0.56}_{-0.79}$ km (CS model) \cite{Raaij2021},
and dashed and dot-dashed ones do 
$R_{1.4M_\odot}=11.94^{+0.76}_{-0.87}$ km \cite{Peter2021} and
$R_{1.4M_\odot}=13.80 \pm 0.47$ km \cite{Brendan2021}, respectively.}
\label{MR2}
\end{center}
\end{figure*}

In Fig.\ref{MR1}, hybrid-star masses are given as a function of radius $R$ 
(left panel) and central baryon density $\rho_{Bc}$ (right panel).
The curves obtained by the Q-EoSs for (a) Q0, (b) Q1 and (c) Q2 are given by
solid curves, and those by the H-EoS for H2 are given by the short-dashed curves. 
The former curves are connected from the latter curves by the hadron-quark phase 
transitions. The maximum masses for (b) Q1 and (c) Q2 are over $2M_\odot$ substantially.
It is noted that the Q-EoS for (a) Q0 derived from $V_{EME}$ is still stiff enough
to reach $2M_\odot$ without help of the repulsive contributions of 
$V_{OGE}$ and $V_{MPP}$. This repulsive components in $V_{EME}$ come from
vector-meson and pomeron exchanges between quarks.
The rectangle in the left panel indicates the region of mass 
$2.072^{+0.067}_{-0.066}$M$_\odot$ and radius $12.39^{+1.30}_{-0.98}$ km
\cite{Riley2021} for the most massive neutron star PSR J0740+6620. 
The $MR$ curves for Q1 and Q2 are found to pass through this rectangle.

In the left panel of Fig.\ref{MR2}, hybrid-star masses are drawn 
as a function of radius $R$, where the Q-EoSs for Q2 and 
H-EoSs for (a) H1, (b) H2 and (c) H3 are used. 
Short-dashed, long-dashed and dot-dashed curves are obtained with
H-EoSs for H1, H2 and H3, respectively.
Solid curves show deviations by transitions from hadronic-matter
to quark-matter phases. 
The maximum masses in the figures are as follows:
In the cases of H-EoSs, they are
$1.82M_\odot$ (H1), $1.94M_\odot$ (H2) and $2.07M_\odot$ (H3).
In the cases of including hadron-quark transitions, they are 
$2.25M_\odot$ (H1+Q2), $2.25M_\odot$ (H2+Q2) and $2.28M_\odot$ (H3+Q2). 
The maximum masses are noted to be determined by the Q-EoSs,
being larger than those given by the H-EoSs.
The rectangle in the left panel is the same as that in Fig.\ref{MR1},
indicating the mass-radius region obtained from the observation \cite{Riley2021}.
The $MR$ curves for the Q-EoSs pass through the rectangle,
though those for the H-EoSs (dashed curves) are below this rectangle.

The radii $R$ at $1.4M_\odot$ ($R_{1.4M_\odot}$) are given as follows:
The values of $R_{1.4M_\odot}$ are 12.5 km (H1+Q2), 13.3 km (H2+Q2) 
and 13.6 km (H3+Q2), being obtained in the cases of 
including hadron-quark transitions. The similar values are obtained 
by using H-EoSs only, which means that the values of 
$R_{1.4M_\odot}$ are determined by H-EoSs.
In the figure,
dotted and solid line segments indicate $R_{1.4M_\odot}=12.33^{+0.76}_{-0.81}$ km
(PP model) and $R_{1.4M_\odot}=12.18^{+0.56}_{-0.79}$ km (CS model) \cite{Raaij2021},
and dashed and dot-dashed ones do 
$R_{1.4M_\odot}=11.94^{+0.76}_{-0.87}$ km \cite{Peter2021} and
$R_{1.4M_\odot}=13.80 \pm 0.47$ km \cite{Brendan2021}, respectively.
The former three line segments (dotted, solid and dashed lines)
are similar with each other, and the $MR$ curve for H1 intersects them.
On the other hand, the $MR$ curves for H2 and H3 intersect the dot-dashed line,
but do not intersect the other three lines. 
In the present stage of the observations for radii of neutron stars,
it is difficult to determine which one of H1, H2 and H3 lead to the most
reasonable EoS.

In the right panel of Fig.\ref{MR2}, hybrid-star masses are drawn as 
a function of central baryon density $\rho_{Bc}$, where the Q-EoS for Q2 is used.
Short-dashed, long-dashed and dot-dashed curves are obtained with
H-EoSs for H1, H2 and H3, respectively. Solid curves show the deviations 
by transitions from hadronic-matter to quark-matter phases. 
In the cases of including hadron-quark transitions, 
the onset values of $\rho_{Bc}$ for quark phases are
0.40 fm$^{-3}$ (H1+Q2), 0.46 fm$^{-3}$ (H2+Q2) and 0.43 fm$^{-3}$ (H3+Q2).

It is useful to compare our results for $MR$ diagrams with those 
in Ref.\cite{Shahrbaf2}, because we employ the method in this
reference for the hadron-quark phase transitions. Though their
quark-matter EoS is based on the nonlocal Nambu-Jona-Lasinio
(nlNJL) model differently from ours, it is found that the quark-phase 
regions of the $MR$ curves in Fig.4 of \cite{Shahrbaf2} are similar 
to ours qualitatively. Especially, maximum masses of 
$2M_{\odot}$ are reproduced well, namely the Q-EoSs are stiff 
similarly in both cases of ours and \cite{Shahrbaf2}.
However, the hadronic-matter regions are different from each other, 
since softer H-EoSs are used in \cite{Shahrbaf2} than ours.

\begin{figure*}[ht]
\begin{center}
\includegraphics*[width=6in,height=3in]{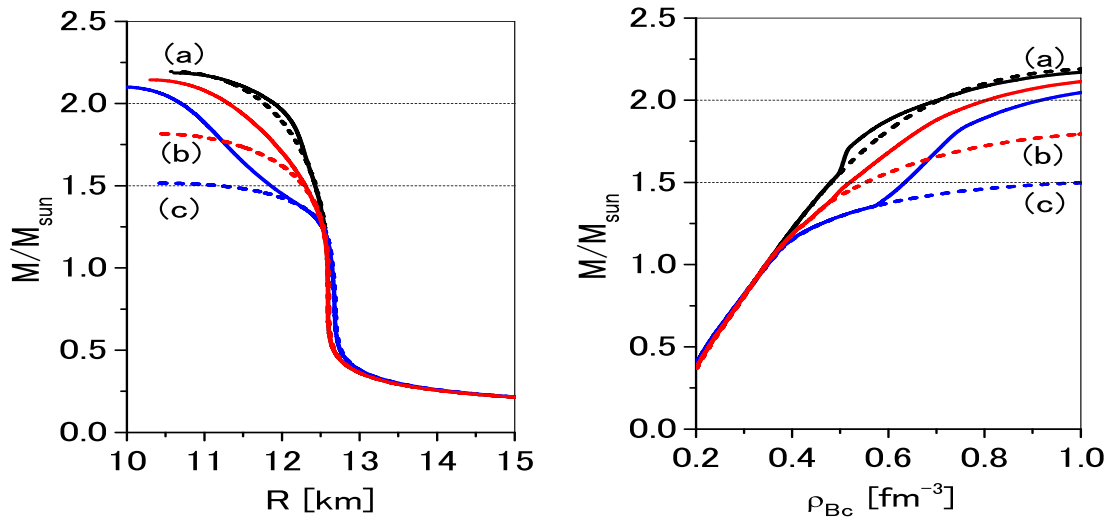}
\caption{hybrid-star masses are given as a function of radius $R$ 
(left panel), and as a function of central density $\rho_{Bc}$ (right panel).
The top dashed curve (a) is obtained from the H-EoS for H0 without hyperons.
The middle dashed curve (b) is from the H-EoS for H1.
The bottom dashed curve (c) is from the H-EoS for H1' including hyperons,
in which the MPP repulsions work only among nucleons.
The solid curves show the deviations by transitions from hadronic phase
to quark-matter phase for Q1.}
\label{MR3}
\end{center}
\end{figure*}

As stated before, the hyperon mixing results in remarkable softening 
of the EoS. In order to avoid this ``hyperon puzzle", the universal repulsions
modeled as MPP are included in our derivations of our H-EoSs.
Here, let us try to use the H-EoS for H1' in which the MPP repulsions work only 
among nucleons. The $BB$ interaction used in \cite{Shahrbaf2} is of this type.
In Fig.\ref{MR3}, hybrid-star masses are given as a function of radius $R$ 
(left panel) and of central density $\rho_{Bc}$ (right panel).
The top dashed curve (a) is obtained from the H-EoS for H0 without hyperons.
The middle dashed curve (b) is from the H-EoS for H1.
The bottom dashed curve (c) is from the H-EoS for H1' including hyperons,
in which the MPP repulsions work only among nucleons.
The solid curves show the deviations by transitions from hadronic phase
to quark-matter phase for Q1.
It should be noted that the large difference from the top dashed curve to
the bottom dashed curve demonstrates the softening of the EoS by hyperon mixing.
Then, the solid curves show the deviations by transitions from hadronic phases
to quark-matter phases for Q1. The lowering of the maximum mass by the EoS
softening turns out to be recovered by the transition to the quark-matter
phase given by the stiff EoS. It is interested that the curves 
for H0+Q1 are similar to those for H0.
The basic feature of the $MR$ curve for H1'+Q1 is similar 
to those of the curves in \cite{Shahrbaf2}.

Our $MR$ diagrams of hybrid stars are derived from H-EoSs for
$BB$ interactions (H1, H2, H3) and Q-EoSs for $QQ$ interactions
(Q0, Q1, Q2): There are nine combinations of H-EoSs and Q-EoSs, 
among which some combinations are used in the above results. 
In Table~\ref{MR}, features of the obtained $MR$ diagrams
in all combinations of (H1, H2, H3) and (Q0, Q1, Q2)
are demonstrated by showing the calculated values of
maximum masses $M_{max}$ and radii $R_{M_{max}}$,
and radii at $1.4M_{\odot}$ ($R_{1.4M_{\odot}}$). 
For comparison, those for H1' and H1'+Q1 are added.
Here, the important features are as follows:
(1) In all cases, the Q-EoSs combined with the H-EoSs are stiff 
enough to reproduce maximum masses over $2M_{\odot}$. 
(2) The values of $R_{1.4M_{\odot}}$ are specified by the H-EoSs.

\begin{table}
\begin{center}
\caption{Maximum masses $M_{max}$ and radii $R_{M_{max}}$,
radii at $1.4M_{\odot}$ $R_{1.4M_{\odot}}$, dimensionless
tidal deformability at $1.4M_{\odot}$ $\Lambda_{1.4M_\odot}$
} 
\label{MR}
\vskip 0.2cm
\begin{tabular}{|l|cccc|}
\hline
 & $M_{max}/M_{\odot}$ & $R_{M_{max}}$ & $R_{1.4M_{\odot}}$  & $\Lambda_{1.4M_{\odot}}$ \\
 &                     &   (km)       &     (km)          &                        \\ 
\hline
H1    & 1.82  &  10.4  & 12.4   & 422   \\
H1+Q0 & 1.99  &  10.0  & 12.4   & 422   \\
H1+Q1 & 2.14  &  10.3  & 12.4   & 422   \\
H1+Q2 & 2.25  &  10.7  & 12.5   & 422   \\
\hline
H1'   & 1.52  &  10.4  & 12.1   & 334   \\
H1'+Q1& 2.10  &  10.0  & 12.2   & 337   \\
\hline
H2    & 1.94  &  10.3  & 13.3   & 671   \\
H2+Q0 & 2.01  &  10.4  & 13.3   & 671   \\
H2+Q1 & 2.16  &  10.6  & 13.3   & 671   \\
H2+Q2 & 2.25  &  10.9  & 13.3   & 671   \\
\hline
H3    & 2.07  &  10.7  & 13.6   & 771   \\
H3+Q0 & 2.04  &  10.7  & 13.6   & 771   \\
H3+Q1 & 2.18  &  10.8  & 13.6   & 771   \\
H3+Q2 & 2.28  &  11.2  & 13.6   & 771   \\
\hline
\end{tabular}
\end{center}
\end{table}

Another constraint for the EoS is given by the tidal deformability,
being the induced quadruple polarizability.
The dimensionless tidal deformability $\Lambda$ is defined as
$\Lambda=(2/3)k_2 (c^2 R/GM)^5$ \cite{Abbott2017}, 
where $c$ is the speed of light, $R$ and $M$ are radius and mass of 
a neutron star and $G$ is the gravitational constant. $k_2$ is the tidal 
Love number describing the response of each star to the external disturbance.
The binary neutron star merger GW170817 give the upper limit on the
tidal deformability of a neutron star with mass $1.4M_\odot$:
$\Lambda_{1.4M_\odot} \leq 800$ \cite{Piekarewitz2019}.
In Table~\ref{MR} are given the calculated values of $\Lambda_{1.4M_\odot}$
for our EoSs, where all values are less than the upper limit of 800.
It should be noted that the values of $\Lambda_{1.4M_\odot}$ are determined
by the H-EoSs, even if the Q-EoSs are combined with them.

\section{Conclusion}
The EoSs and $MR$ diagrams of hybrid stars are obtained
on the basis of our $QQ$ interaction model composed of
the extended meson exchange potential ($V_{EME}$),
the multi-pomeron exchange potential ($V_{MPP}$), 
the instanton exchange potential ($V_{INS}$) and 
the one gluon exchange potential ($V_{OGE}$), whose strengths
are determined on the basis of terrestrial data with
no adhoc parameter to stiffen EoSs.
The repulsive nature of our $QQ$ interaction in high density
region are basically given by $V_{EME}$ including strongly 
repulsive components owing to vector-meson and pomeron exchanges.
Additional repulsions (attractions) are given by
$V_{MPP}$ and $V_{OGE}$ ($V_{INS}$).
The resultant repulsions included in our $QQ$ interaction are 
so strong that the quark-matter EoSs become stiff enough 
to give maximum masses of hybrid stars over $2M_\odot$.

Hadronic-matter EoSs (H-EoS) and quark-matter EoSs (Q-EoS) 
are derived in the same framework based on the BBG theory. 
In quark matter, density-dependent quark masses are introduced
phenomenologically, playing a decisive role in the occurrence 
of phase transition.
Parameters of density dependences are taken so that 
hadron-quark phase transitions are able to occur at reasonable
density region owing to the reduction of the quark masses and 
chemical potentials in quark matter. Our resulting density 
dependence of effective quark mass is similar to the Brown-Rho scaling. 

Our H-EoSs are still not stiff enough to give maximum masses of neutron
stars over $2M_\odot$ due to the softening by hyperon mixing, although 
the stiffness is recovered substantially by universal many-body repulsions. 
In the case of using our $QQ$ interaction model, the Q-EoSs are stiffer
than the H-EoSs and $MR$ curves of hybrid stars shift above those of 
stars with hadronic phases only.
The maximum masses of the formers including quark phases become larger 
than those of the latters, and $MR$ curves are characterized by Q-EoSs
in the mass region higher than about $1.5M_\odot$.
Our Q-EoSs for Q1 and Q2 are stiff enough to give a maximum 
mass over $2M_\odot$. The derived mass and radius are consistent
with the recent measurement for the most massive neutron star 
PSR J0740+6620, obtained by the combining analysis for
the NICER and the other multimessenger data.

In our approach, star radii $R_{1.4M_\odot}$ given by hadronic-matter
EoSs do not changed by the hadron-quark phase transitions, namely 
they are determined by H-EoSs regardless of Q-EoSs.
There are three estimates of $R_{1.4M_\odot}$ based on
the NICER measurements and the other multimessenger data.
Two of them give $R_{1.4M_\odot}=11.1 - 13.1$ km,
and the other $R_{1.4M_\odot}=13.1 - 14.4$ km.
Our H-EoS for H1 (H2 or H3) is consistent with the former (latter). 

Our H-EoSs and Q-EoSs lead to $MR$ diagrams of hybrid stars 
consistent with the recent observations for masses and radii.  

\newpage

\section*{Acknowledgments}
{The authors would like to thank D. Blaschke 
for valuable comments and fruitful discussions. 
This work was supported by JSPS KAKENHI (No.20K03951 and No.20H04742).}

\end{document}